\documentclass[10pt,a4paper]{article}

\usepackage{amsmath}
\usepackage{amsfonts}
\usepackage{latexsym}
\usepackage{psfig}
\usepackage{graphicx}

\begin{document}

\title{\Large Transitive Text Mining for Information\protect\linebreak Extraction and Hypothesis Generation}
\author{Johannes Stegmann\footnote{Medical Library, corresponding author: johannes.stegmann@charite.de.} \hspace{0.1cm}and Guenter Grohmann\footnote{Institute of Medical Informatics, Biometry and Epidemiology}
 \\
 Charit\'e - Campus Benjamin Franklin, Berlin, Germany\\
 }
\maketitle

\begin{abstract}
Transitive text mining - also named Swanson Linking (SL) after its primary and principal researcher - tries to establish 
meaningful links between literature sets which are virtually disjoint in the sense that each does not mention the main 
concept of the other. If successful, SL may give rise to the development of new hypotheses.

In this communication we describe our approach to transitive text mining which employs co-occurrence analysis of the 
medical subject headings (MeSH), the descriptors assigned to papers indexed in PubMed. In addition, we will outline the 
current state of our web-based information system which will enable our users to perform literature-driven hypothesis 
building on their own.\\

\end{abstract}

\section{Introduction} \indent
\indent Transitive text mining tries to link the major themes of disjoint literature sets. Don Swanson was the first to describe 
the disclosure of  ''hidden'', i.e. unpublished but implicit links between concepts not mentioning each other in their 
respective literature representations (Swanson (1986, 1988, 1991)). Later, the principle of his method was termed 
{\em Swanson Linking} ({\em SL}) which may be defined as finding disjoint literature partners by establishing meaningful links between 
them using information retrieval from bibliographic databases (Stegmann and Grohmann(2003)).

The published examples of SL involve basically three different sets of literature: (i) a problem-based literature - e.g. 
describing a disease - is referred to as ''source''; (ii) a literature not being mentioned in the source literature but 
possibly contributing to problem solving is called ''target''; (iii) a literature representing a major concept which is 
relevant for and occurs in both, source and target literature, is labeled ''intermediate''(Swanson and Smalheiser (1999)). 
The discovery process might normally proceed from source to target via intermediate; however, the reverse order is 
naturally conceivable, and any coherent literature set regarded as ''intermediate'' may be explored for source and target concepts simultaneously.

Different approaches have been developed to detect in the investigated literature sets key terms representing possible 
intermediate and target or source concepts. Some authors extract title and abstract words and phrases (Swanson and 
Smalheiser (1999), Gordon and Lindsay (1996), Gordon and Dumais (1998), Lindsay and Gordon (1999), Weeber et al. (2001)) 
or extract the medical subject headings (MeSH) assigned to documents indexed in PubMed (Srinivasan (2004)) and try to 
find relevant terms on top of ranked lists. We use the co-word analysis technique described by Callon et al. (1991) for 
clustering of extracted MeSH terms and for two-dimensional visualisation of the clusters according to their internal 
density and external centrality in so-called ''strategical diagrams''. We found that in some cases key concepts relevant 
to the discovery process can be identified on the basis of some positional and/or numerical characteristics of the 
respective clusters (Stegmann and Grohmann (2003), Stegmann and Grohmann (2004)). 

In this communication we discuss and value these features with the following examples of transitive text mining: 
{\em Raynaud's Disease - Fish Oil} (Swanson (1986)), {\em Migraine - Magnesium} (Swanson (1988)), 
{\em Schizophrenia - Phospholipase $A_{2}$} (Smalheiser and Swanson (1998)).

In addition, we shortly describe our web-based information system which will enable our users to analyse the knowledge
domain represented by a literature set and to perform transitive text mining on their own. 

\section{Methods} \indent
\indent PubMed searches were performed as indicated at the legends. The retrieved document sets were downloaded in MEDLINE format.
Extraction of MeSH terms, subsequent co-occurrence analysis, term clustering and calculation of cluster density and 
centrality were performed as described (Stegmann and Grohmann (2003), Stegmann and Grohmann (2004)). A brief description of
the cluster process follows: co-occurrence strength of MeSH term pairs was calculated as  {\em Equivalence Index} (Callon et al. (1991))
\begin{displaymath}
  E_{ij} = \frac{ C^{2}_{ij} }{ C_{i} C_{j} },
\end{displaymath}
where \(C_{ij}\) is the number of co-occurrences of terms i and j (i.e. the number of documents in which terms i and j co-occur),
and \(C_{i}\) and \(C_{j}\) are the number of occurrences of term i and j, respectively. Multiple occurrences of a MeSH term 
within the MeSH fields of a document (e.g. with different subheadings) are ignored. A threshold of \(E_{ij} \geq 0.05\) was
applied. The cluster process starts with the term pair exhibiting the highest equivalence index. Of those remaining terms 
having links with the cluster members the term with the highest link strength is added to the cluster. Cluster size is 
limited to 3 - 10 terms. The clusters of a literature set are graphically displayed according to their mean internal link
strength (density) and the sum of their external link strength (centrality).

The tools for MeSH term extraction, calculation of equivalence indices, cluster generation and calculation of cluster 
density and centrality have been programmed in PERL and JAVA. The JAVA programs are part of our web-based information 
system {\em Charit\'e-Mlink}. Cluster diagrams were created using the software package R (R Development Core Team (2004)). 

\section{Results and Discussion} \indent
\indent The strategical diagrams have a two-fold function: (i) they should allow the identification of clusters containing
terms of potential interest for the transitive discovery process; (ii) they represent knowledge domains (as far as they are
comprised by the database retrieval) which can be analysed in terms of centrality indicating the importance of clusters 
and cluster terms for the whole domain, and in terms of density indicating the strength of the local coherence of (sub-) themes
expressed by the cluster terms (Callon et al. (1991)).

Figure \ref{fig:rd}, \ref{fig:mi}, \ref{fig:sc} show strategical diagrams of {\em source} literatures which allow the identification of {\em intermediate} 
terms which in turn are prerequisites for the detection of {\em target} concepts. Diagrams of {\em intermediate} 
literatures are displayed in Figure \ref{fig:bv}, \ref{fig:sd}, \ref{fig:pa}. They harbor both, {\em target} and {\em source} terms.

\subsection{Source literature} 
\begin{figure}
  \centerline{
  \includegraphics{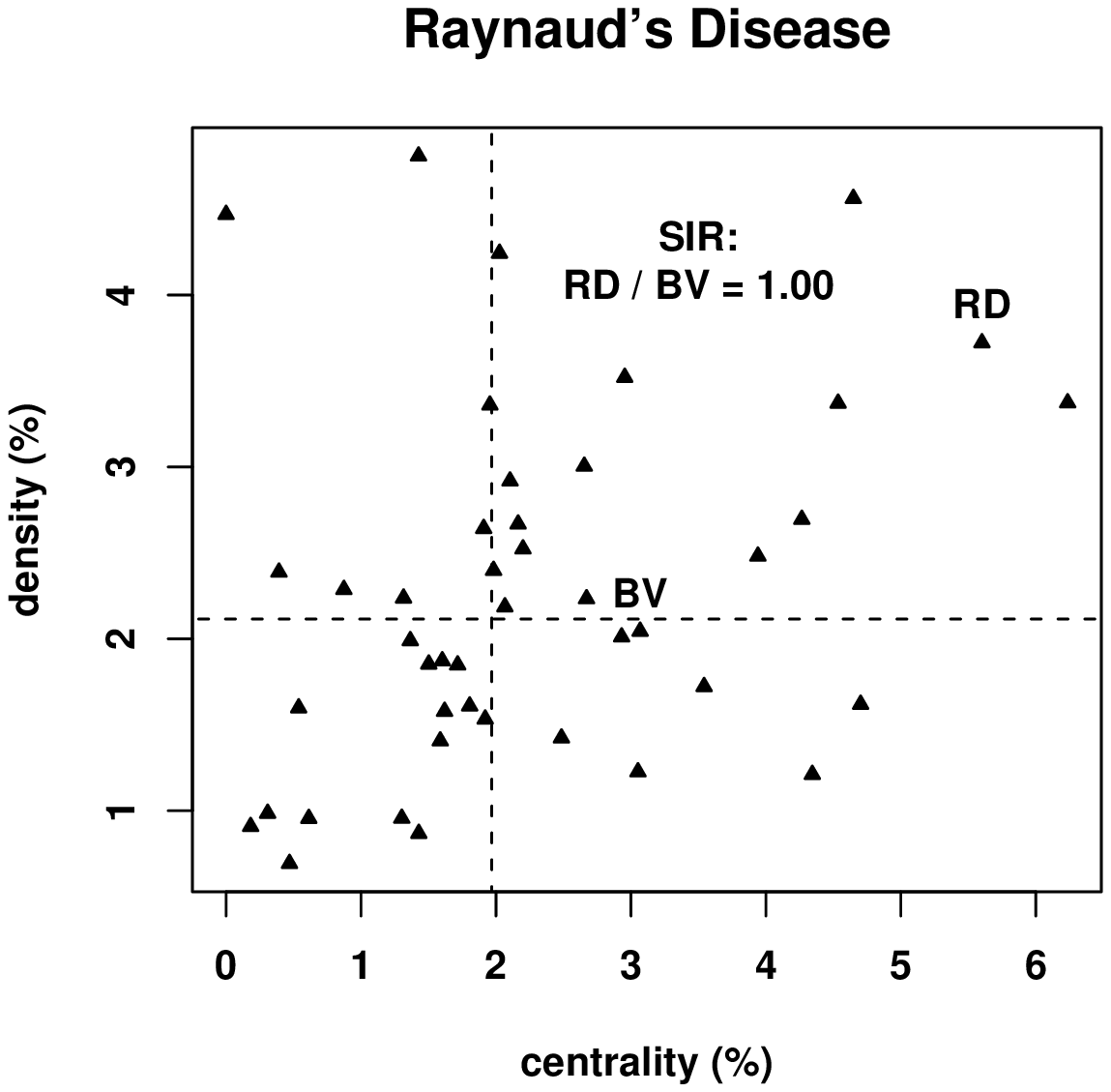}}
    \caption{Strategical diagram of the {\em Raynaud's Disease\textsuperscript{\ddag}} literature set.\newline 
        \textsuperscript{\ddag}PubMed title search for ''raynaud*'', publication years 1966-1985.\newline
         No. of documents: 802, no. of distinct MeSH terms: 454, no. of clusters: 44.\newline
         Triangles: indicate cluster positions, dotted lines: indicate medians.\newline             
         RD: cluster containing the source term {\em Raynaud's Disease}.\newline
         BV: cluster containing the intermediate term {\em Blood Viscosity}.\newline
         SIR: source-intermediate ratio.}
  \label{fig:rd}
\end{figure}

\begin{figure}
  \centerline{
    \includegraphics{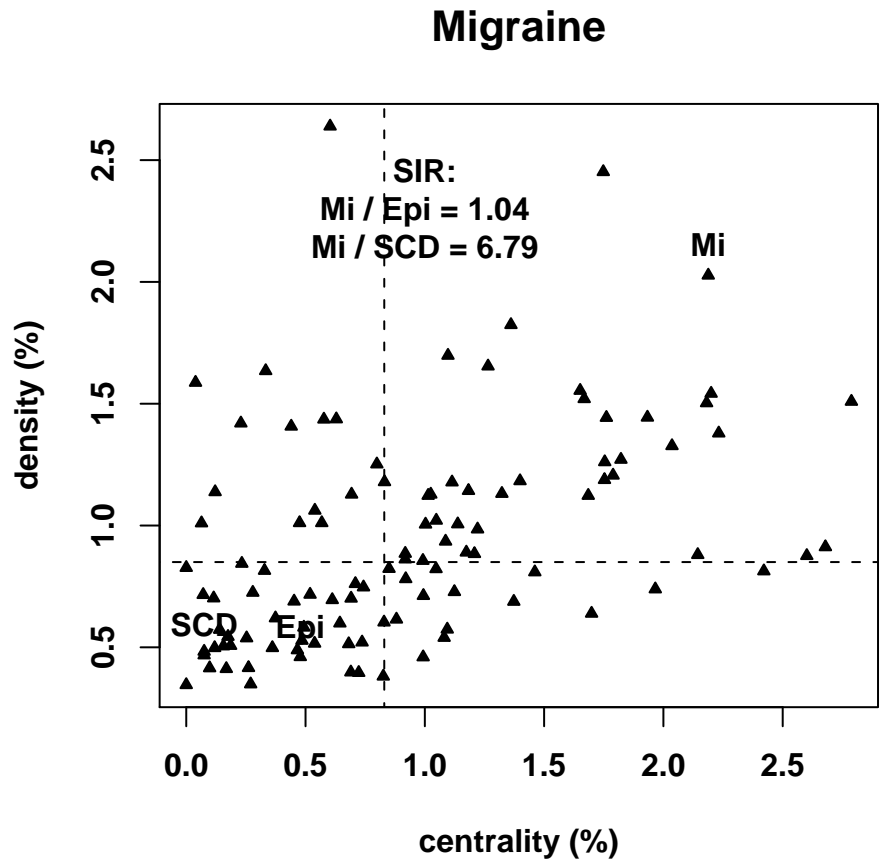}}
    \caption{Strategical diagram of the {\em Migraine\textsuperscript{\ddag}} literature set.\newline 
        \textsuperscript{\ddag}PubMed title search for ''migraine'', publication years 1966-1987.\newline
         No. of documents: 2583, no. of distinct MeSH terms: 1021, no. of clusters: 106.\newline
         Mi: cluster containing the source term {\em Migraine}.\newline
         Epi: cluster containing the intermediate term {\em Epilepsy}.\newline
         SCD: cluster containing the intermediate term {\em Spreading Cortical Depression}.\newline
         Other details: see Figure \ref{fig:rd}. }  
  \label{fig:mi}
\end{figure}

\begin{figure}
  \centerline{
    \includegraphics{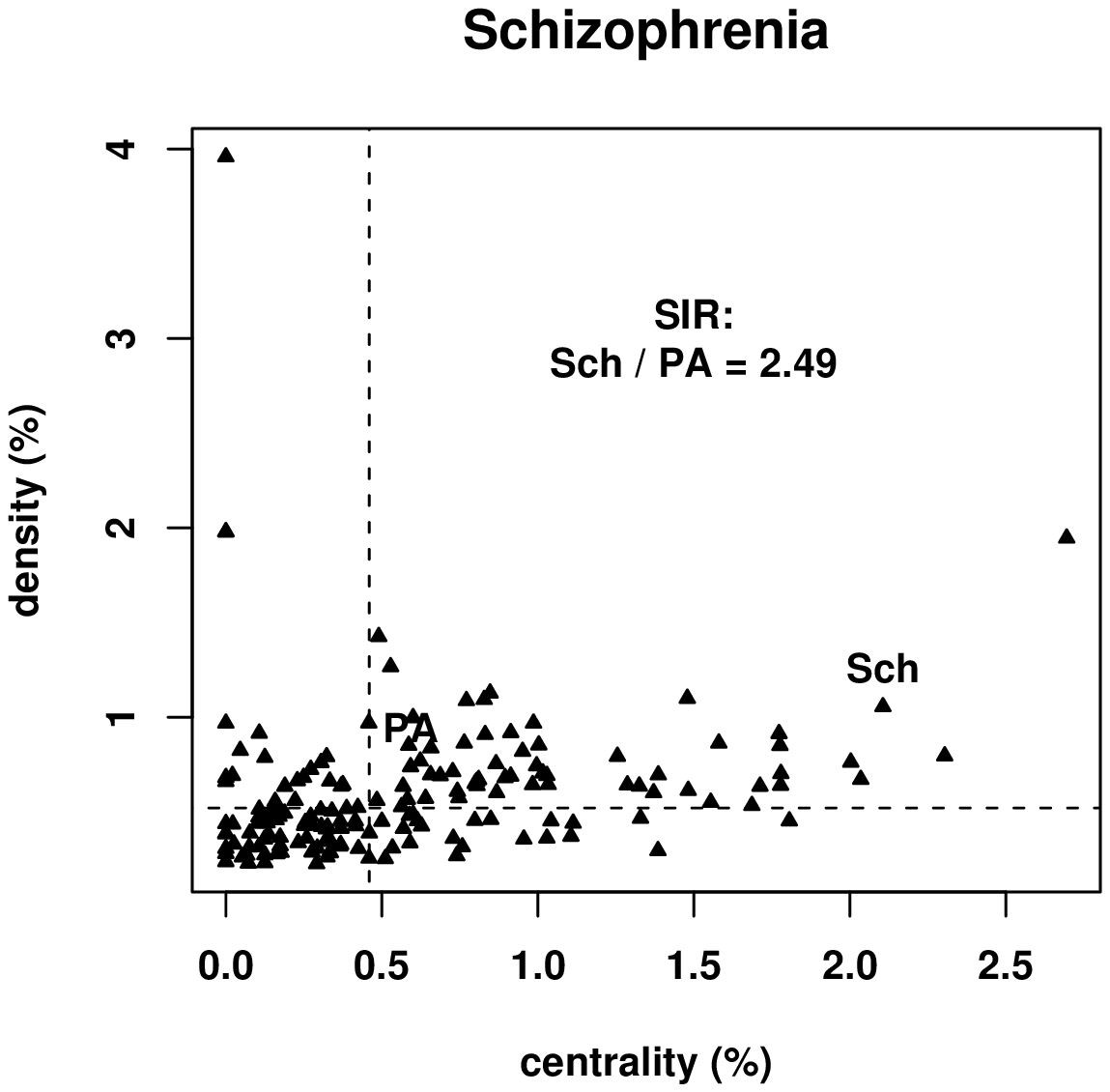}}
    \caption{Strategical diagram of the {\em Schizophrenia\textsuperscript{\ddag}} literature set.\newline 
        \textsuperscript{\ddag}PubMed title search for ''schizophrenia'', publication years 1966-1985.\newline
         No. of documents: 6225, no. of distinct MeSH terms: 1598, no. of clusters: 164.\newline
         Sch: cluster containing the source term {\em Schizophrenia}.\newline
         PA: cluster containing the intermediate term {\em Platelet Aggregation}.\newline
         Other details: see Figure \ref{fig:rd}. }  
  \label{fig:sc}
\end{figure}

\indent
\indent In the diagrams of source literatures the clusters with the terms defining the literature are
located in regions of high centrality and density, as one can expect (Figure \ref{fig:rd}, \ref{fig:mi}, \ref{fig:sc}).
The clusters containing some of the already known (from Swanson's work) intermediate terms are indicated 
(Figure \ref{fig:rd}, \ref{fig:mi}). The terms {\em Spreading Cortical Depression} and {\em Epilepsy}, being intermediate 
for the {\em Migraine - Magnesium} literature track occur in clusters located in the below-median centrality and density 
region of the diagram, that is in the periphery of the knowledge domain "Migraine" (Figure \ref{fig:mi}). In contrast, 
the cluster containing the term {\em Blood Viscosity} as an intermediate for the {\em Raynaud's Disease - Fish Oil} 
literatures has a higher centrality and about median density (Figure \ref{fig:rd}). For each cluster with a 
\(centrality > 0\) a {\em centrality/density ratio} ({\em cdr}) can be calculated as the quotient of its centrality and 
density. Dividing the cdr of the source cluster by the cdr of an intermediate cluster gives a 
{\em Source-Intermediate Ratio} ({\em SIR)}. SIR is around \(1\) for the intermediate terms {\em Blood Viscosity} 
(Figure \ref{fig:rd}) and {\em Epilepsy} (Figure \ref{fig:mi}). However, the SIR of the intermediate 
{\em Spreading Cortical Depression} (Figure \ref{fig:mi}) is quite different from \(1\). 
The analysis of other source literatures (not shown) also identifies some intermediate terms in clusters being located
in regions of low density and centrality and/or showing a SIR of around \(1\) so that these characteristics may be taken
as indicators where to start the screening of the clusters for terms of possible relevance for the discovery process.
However, the other clusters should be screened, too. Due to the cluster method used the members of a cluster show some
similarity to each other and oftenly define the different aspects of a more general theme which may be helpful in generating
ideas of intermediate concepts. The diagram of the {\em Schizophrenia} source literature (Figure \ref{fig:sc}) may serve as 
an example: here, the intermediate cluster is neither located below the medians nor shows a SIR of around \(1\). In fact, 
we identified it tentatively as an intermediate because it contains the term {\em Platelet Aggregation} which is also
an intermediate term in the {\em Raynaud's Disease} and {\em Migraine} literature (not shown) and because that term 
represents some physiological property. It is generally a good idea to look for candidate intermediate terms dealing
with physiological conditions (Weeber et al. (2001)).

\subsection{Intermediate literature} 
\begin{figure}
  \centerline{
    \includegraphics{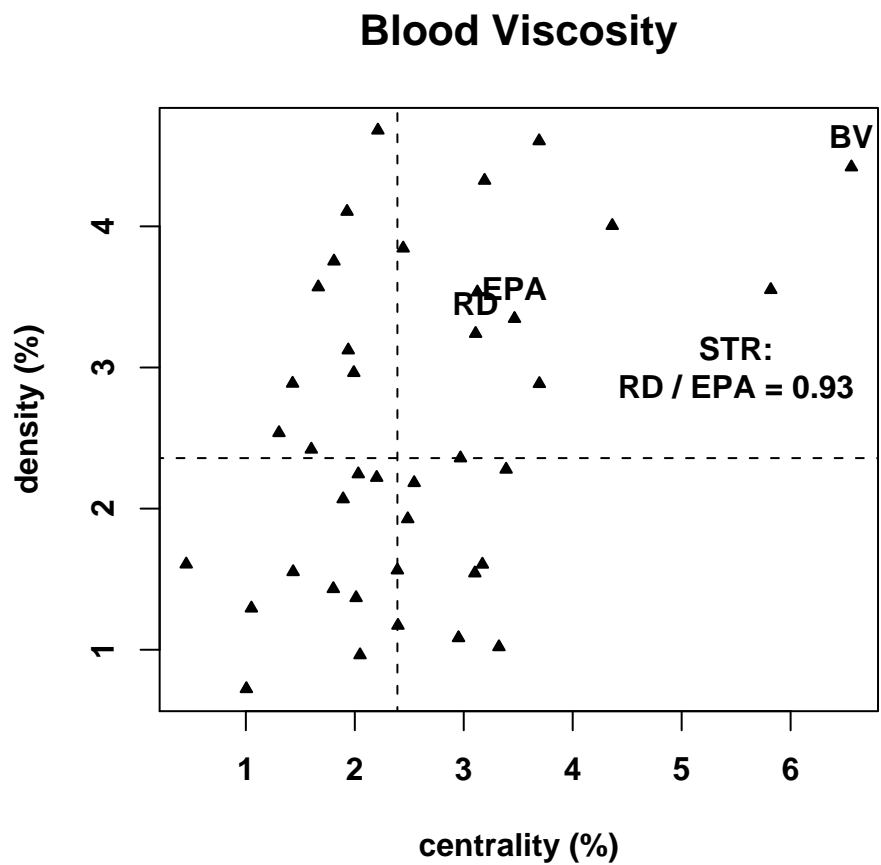}}
    \caption{Strategical diagram of the {\em Blood Viscosity\textsuperscript{\ddag}} literature set.\newline 
        \textsuperscript{\ddag}PubMed title search for ''blood viscosity'', publication years 1966-1987.\newline
         No. of documents: 326, no. distinct of MeSH terms: 293, no. of clusters: 39.\newline
         BV: cluster containing the intermediate term {\em Blood Viscosity}.\newline
         RD: cluster containing the source term {\em Raynaud's Disease}.\newline
         EPA: cluster containing the target terms {\em Eicosapentaenoic Acid} and {\em Fish Oils}.\newline
         STR: source-target ratio.\newline
         Other details: see Figure \ref{fig:rd}. }
 \label{fig:bv}
\end{figure}

\begin{figure}
  \centerline{
    \includegraphics{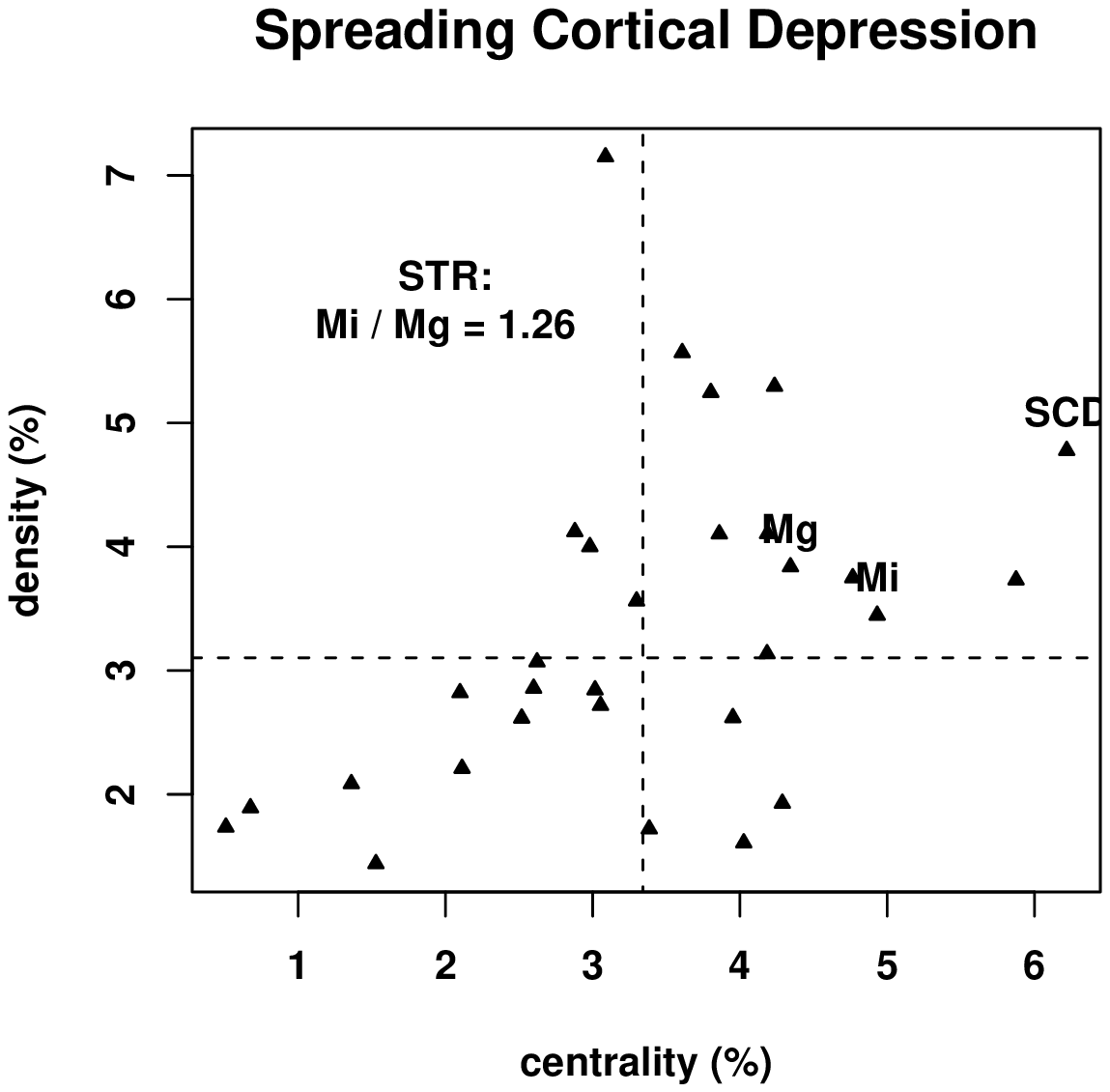}}
    \caption{Strategical diagram of the {\em Spreading Cortical Depression\textsuperscript{\ddag}} literature set.\newline 
        \textsuperscript{\ddag}PubMed title search for ''spreading cortical depression'', publication years 1966-1985.\newline
         No. of documents: 502, no. distinct of MeSH terms: 391, no. of clusters: 30.\newline
         SCD: cluster containing the intermediate term {\em Spreading Cortical Depression}.\newline
         Mi: cluster containing the source term {\em Migraine}.\newline
         Mg: cluster containing the target terms {\em Magnesium}.\newline
         Other details: see Figure \ref{fig:rd} and Figure \ref{fig:bv}. }
 \label{fig:sd}
\end{figure}

\begin{figure}
  \centerline{
    \includegraphics{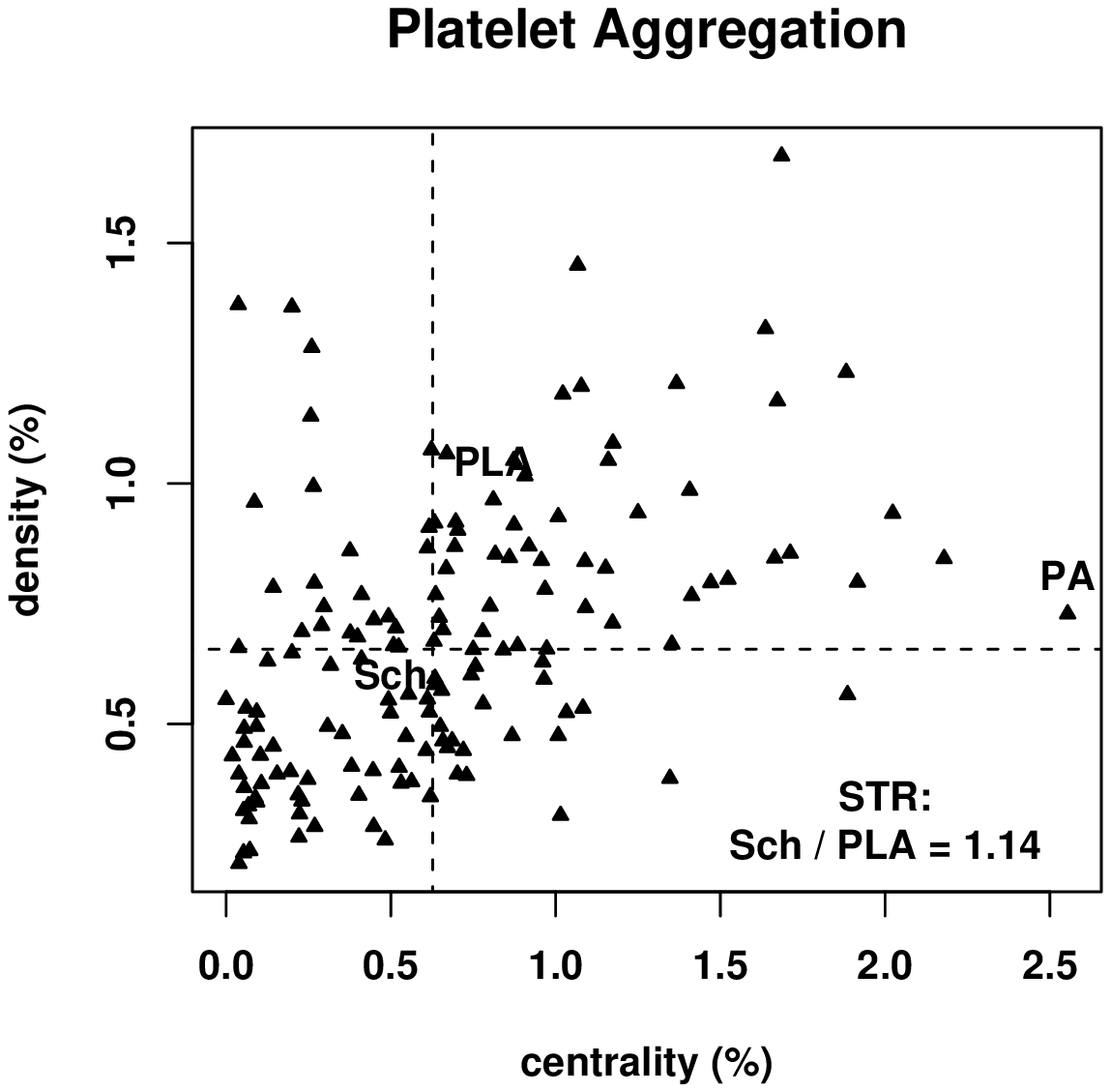}}
    \caption{Strategical diagram of the {\em Platelet Aggregation\textsuperscript{\ddag}} literature set.\newline 
        \textsuperscript{\ddag}PubMed title search for ''platelet aggregation'', publication years 1966-1985.\newline
         No. of documents: 2638, no. distinct of MeSH terms: 1449, no. of clusters: 148.\newline
         PA: cluster containing the intermediate term {\em Platelet Aggregation}.\newline
         Sch: cluster containing the source term {\em Schizophrenia}.\newline
         PLA: cluster containing the target terms {\em Phospholipase A}.\newline
         Other details: see Figure \ref{fig:rd} and Figure \ref{fig:bv}. }
 \label{fig:pa}
\end{figure}

\indent 
\indent Figure \ref{fig:bv}, \ref{fig:sd}, \ref{fig:pa} show the strategical diagrams of the intermediate literatures represented
by the intermediate term identified in Figure \ref{fig:rd}, \ref{fig:mi}, \ref{fig:sc}. As to be expected, the clusters 
containing the main concept of the literature sets show high centrality. The intermediate literatures contain by definition
the respective source terms because the former was choosen due to the identification of its main term from the diagrams of the
source literatures. Now, in the diagrams of the intermediate literatures the clusters containing a source term may be 
regarded as a guide to possibel target concepts. In the {\em Blood Viscosity} literature diagram (Figure \ref{fig:bv}) 
the cluster containig the target terms {\em Eicosapentaenoic Acid} and {\em Fish Oils} are not only located close together
but also show similiar centraliy/density ratios which give a quotient ({\em Source-Target Ratio}, {\em STR}) of
about \(\)1. In the diagram of the intermediate {\em Spreading Cortical Depression} literature set (Figure \ref{fig:sd}) we also find source and
target clusters in close vicinty, the STR, however is well above 1. In the diagram of the intermediate 
{\em Platelet Aggregation} literature (Figure \ref{fig:pa}) source and target clusters are not so close together, but the STR value is around
\(1\). All target terms are already known by Swanson's work.

As with the source literature we can start the screening of intermediate literature clusters exhibiting similiar 
centrality/density ratios and/or being in the neighbourhood of source clusters. However, we must also say that some
target terms are found in clusters outside of this frame (not shown); the screening of other clusters is always advisable.
In addition, dealing with large literature sets consisting of many thousand documents very many clusters have to be screened.
We did not yet experiment with variable cluster sizes but higher number of terms per cluster might affect the cluster
readability. Thus, other text mining methods should be employed. For example, Gordon and Dumais (1998) applied 
{\em Latent Semantic Analysis} ({\em LSA}) to the analysis of the {\em Raynaud's Disease} literature and found relevant 
intermediate terms at high ranks on list generated from title and abstract words. They failed, however, to find target 
terms at equally high positions on lists derived from the intermediate literature after LSA treatment. 
We are currently investigating the usefulness of LSA for the analysis of document-by-MeSH-term matrices (in preparation).

\subsection{Charit\'e-Mlink} \indent
\indent Our web-based information system {\em Charit\'e-Mlink} enables the user to upload PubMed literature sets and to navigate 
in the information space constituted by the MeSH term clusters generated by the system. In addition, the system makes
suggestions with respect to clusters containing terms of potential relevance to a discovery process based on transitive 
text mining as described in the previous sections. The first version of {\em Charit\'e-Mlink} has been released in 
August 2005 and can be accessed at http://mlink.charite.de/.

\section{Conclusion} \indent
\indent We have described an approach to transitive text mining which is based on the co-occurrence analysis and subsequent
clustering of the MeSH terms assigned to PubMed documents. Our results allow some suggestions which clusters should
be screened at first in a discovery process. Future work is necessary employing other text mining methods in order to
generate hypotheses which cannot derived from one knowledge domain only.\\\\

\hspace{3.0cm} \textbf{Acknowledgements}\\\\
\indent This paper has been presented in part at the 29\textsuperscript{th}Annual Conference of the German Classification Society, 
Magdeburg, March 9-11, 2005.
 
Our work is currently supported by the Deutsche Forschungsgemeinschaft, grant no.
 LIS 4 - 542 81.

\end{document}